# Revealing localised dark-exciton populations in 2D perovskites via magneto-optical microscopy

Christopher G. Bailey,[*,1] Adrian Mena,[2] Tik Lun Leung,[1] Nicholas P. Sloane,[2] Chwenhaw Liao,[1] David R. McKenzie,[1] Dane R. McCamey,[*,2] and Anita Ho-Baillie[*,1]

[1]*School of Physics, University of Sydney, Sydney, Australia*

[2]*ARC Centre of Excellence in Exciton Science, School of Physics, University of New South Wales, Sydney, Australia*

\* E-mail: Christopher.Bailey@sydney.edu.au; Dane.McCamey@unsw.edu.au; Anita.Ho-Baillie@sydney.edu.au

## Abstract

The successful development of optoelectronic devices is contingent on a detailed understanding of interactions between light and excited energy states in photoactive materials. In 2D perovskites, excitons are the dominant photogenerated species and their energetic structure plays a pivotal role, governing photon absorption and emission processes. In these materials, dark exciton states can undergo photoluminescence due to the relaxation of selection rules and this process can be modulated by an external magnetic field, enabling unambiguous identification of the exciton fine structure. Previous reports of magneto-optical spectroscopy on 2D perovskites have been restricted to the macroscopic response, where key information is lost regarding the microscopic heterogeneity of the photoluminescence. Here, we use magneto-optical microscopy for the first time on perovskite materials to elucidate the spatial variation of exciton emission processes. In 2D perovskite thin films, we distinguish between regions of localised bright and dark exciton populations, correlated to the film morphology. In single crystals, we show that dark excitons become localised at the edges, where excitons can be trapped in two distinct types of sub-gap states. This work represents significant progress in understanding the properties of exciton emission in 2D perovskites, which is crucial for the development of optoelectronic technology.

## Introduction

Research on 2D metal halide perovskites has provided a leap in the progress of next-generation optoelectronics [1–10] owing to their unique advantageous properties. [4–6] 'Pure-2D' perovskites consist of an alternating structure where single metal-halide octahedral layers ($n$ = 1) are separated by a layer of spacer cations, which are typically organic molecules. These, in addition to 'quasi-2D' perovskites with multiple ($n$ > 1) octahedral layers separated by spacer cations, have been utilised in high-efficiency solar cells[1–6,11,12] and other types of optoelectronic devices,[7–10] improving their performance and stability. The improved ambient stability compared to 3D perovskites has been ascribed to factors such as the hydrophobicity of the organic spacer cation,[13–15] inhibition of ion migration,[13,16,17] and molecular interactions of spacer cations.[18] Stability is essential for developing mature solar cell technology, meeting industry standards,[19–22] and passing testing protocols[23] before successful commercialisation.

The exciton binding energy in 2D perovskites is typically larger than those in other forms of perovskite, including bulk 3D perovskites and even nanocrystals.[10,24–29] This is due to both the large mismatch in dielectric constant and the quantum confinement between alternating planes which resemble quantum well heterostructures, causing carriers to be strongly bound within their respective plane. The presence of excitons at room temperature provides an opportunity to utilise 2D perovskites for light-emitting applications[30–33] and as a testbed for studying exciton-related phenomena including polarons,[34,35] self-trapping,[36,37] polaritons,[38–40] Bose–Einstein condensation,[39] and spin-dependent processes.[41–45] Despite technological advancements, the underlying science pertaining to 2D perovskites requires further research effort due to many contentious reports in the literature which generally lack consensus.

Developing an accurate picture of photoexcited energy states in 2D perovskites is crucial for light-harvesting and emitting applications since they dictate photon absorption and carrier recombination mechanisms. Therefore, engineering of these energy states is an important pathway to optimising optoelectronic devices. For example, the coupling of light to an exciton state depends on factors such as its angular momentum, resulting in 'bright' and 'dark' excitons. In Ruddlesden–Popper (RP) phase 2D perovskites, 'dark' exciton states can emit photons, despite the process violating typical selection rules that otherwise prevent electronic transitions.[46] Additionally, dark excitons are spin-polarised species with long lifetimes found in various material systems[46–48] and hence are potential qubit candidates for quantum computing.[49,50]

Magneto-optical spectroscopy has proved to be an excellent tool for understanding the exciton fine structure of 2D perovskites.[41,51–58] Mixing between dark and bright exciton eigenstates can be modified by an externally applied magnetic field, allowing for unambiguous determination of the exciton fine structure via optical spectroscopy.[41,42,51,52,54,59] However, prior studies have only investigated magnetic field effects as a macroscopic bulk property of 2D perovskites, without consideration of any potential microscopic variation. [41,42,51,52,54,59]

Here, we use low-temperature magneto-optical microscopy to probe exciton emission for the first time on any perovskite material. The spatial information reveals localised exciton populations in the RP phase 2D perovskite phenethylammonium lead iodide ((PEA)$_2$PbI$_4$). Our findings uncover the effect of observed inherent spatial heterogeneity of 2D perovskites on light-emitting properties. The results of this work are imperative for the development of next-generation optoelectronics and exploring new applications which harness the selectivity of spin states or interaction with an external magnetic field, enabling sensing, spintronics, and quantum computing technologies.

## Results and Discussion

### Magneto-optical imaging of (PEA)$_2$PbI$_4$ thin films

We first investigate spatial variation of low-temperature photoluminescence (PL) in (PEA)$_2$PbI$_4$ thin films (see Experimental Section for details of film preparation) using the experimental setup illustrated in Figure 1a. The PL from the sample can be either spatially or spectrally resolved under a variable external magnetic field, applied in the plane parallel to substrate and perpendicular to optical path (Voigt geometry). Figures 1b and c show the principle of dark-exciton brightening in 2D perovskites under a magnetic field in the Voigt geometry, described in more detail below.[41,42,51,52,54,59]

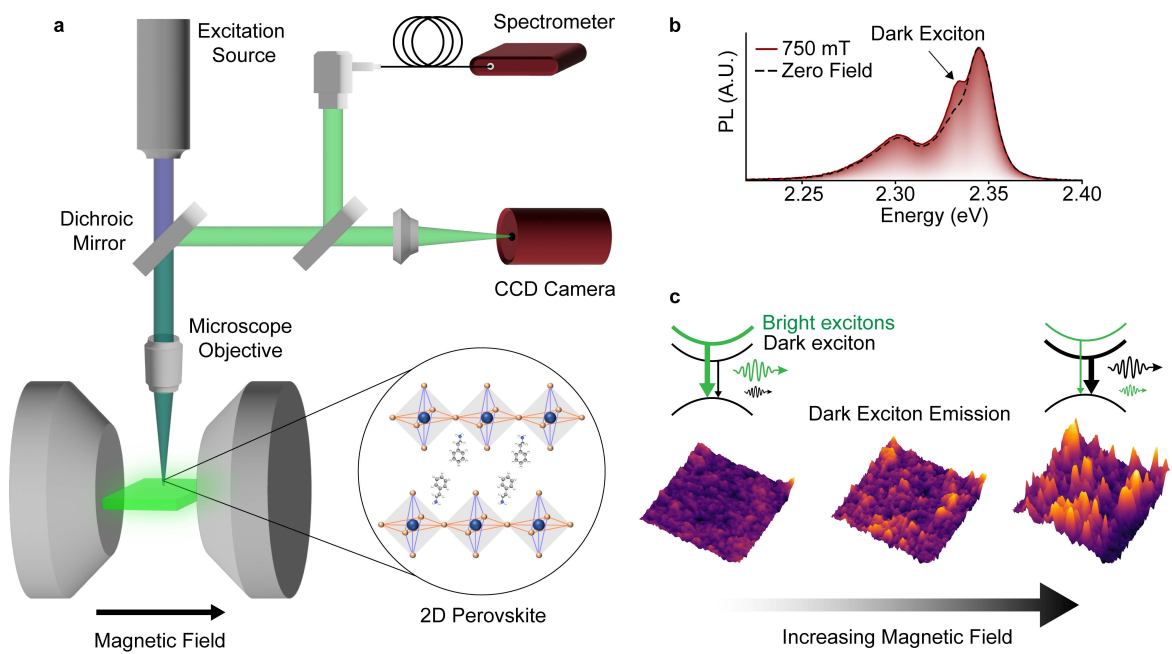

Figure 1: **Modulation of dark-exciton emission using an applied in-plane magnetic field.** a) Schematic of low-temperature PL microscopy on 2D perovskites under an applied in-plane magnetic field. b) PL spectrum of a (PEA)$_2$PbI$_4$ thin film, showing the increased dark-exciton emission under magnetic field. c) Schematic illustrating principle of dark-exciton brightening in a 2D perovskite thin film.

Firstly, we obtained a PL image of the (PEA)$_2$PbI$_4$ thin film at 3 K without an applied magnetic field, shown in Figure 2a, revealing non-uniform PL with localised regions of brighter or darker emission compared to the average intensity. Using bright- and dark-field optical microscopies, and scanning electron microscopy (SEM), we relate the variation of PL to film morphology (Figures S1, S2 and Supporting Note 1). Specifically, we observe high PL intensity at grain boundaries, likely due to a waveguiding effect caused by the two-dimensional confinement of photons in the plate-like grains, similar to that observed in thin single-crystal flakes of (PEA)$_2$PbI$_4$. [60,61]

The PL spectrum of the (PEA)$_2$PbI$_4$ thin film consists of features for bright excitons (BX), dark excitons (DX), and lower energy excitons (LX) (Figure 1b). The BX transition is a triplet state with angular momentum, $J = 1$, comprised of excitons with dipole moments oriented in the three crystallographic directions, with two in-plane ($x$, $y$) and one out-of-plane ($z$).[42,59,62] We represent BX as a single transition for simplicity and find that the observed spectra can be fitted with the sum of three Gaussian functions representing BX, DX, and LX (Figure 2d), using the method described previously.[51] Illustrated in Figure 1a–c, an applied magnetic field at low temperature in the Voigt geometry causes DX emission to be enhanced, while emission from BX is reduced due to mixing of DX with the in-plane BX sub-levels (Figure 2e).[41,42,51]

We repeated the PL imaging experiment under an external magnetic field allowing us to probe the variation of DX brightening across the film (Figure 2b). We can then calculate the relative PL intensity change under an applied magnetic field as,

$$\text{PL Intensity Change (\%)} = \frac{\Delta \text{PL}}{\text{PL}} \times 100 = \frac{\text{PL}(B) - \text{PL}(0)}{\text{PL}(0)} \times 100, \tag{1}$$

where PL(0) is the PL at zero magnetic field, and PL($B$) is the PL under an applied magnetic field, $B$. We refer to the PL intensity change ($\Delta$PL/PL) as $\Delta$PL for brevity.

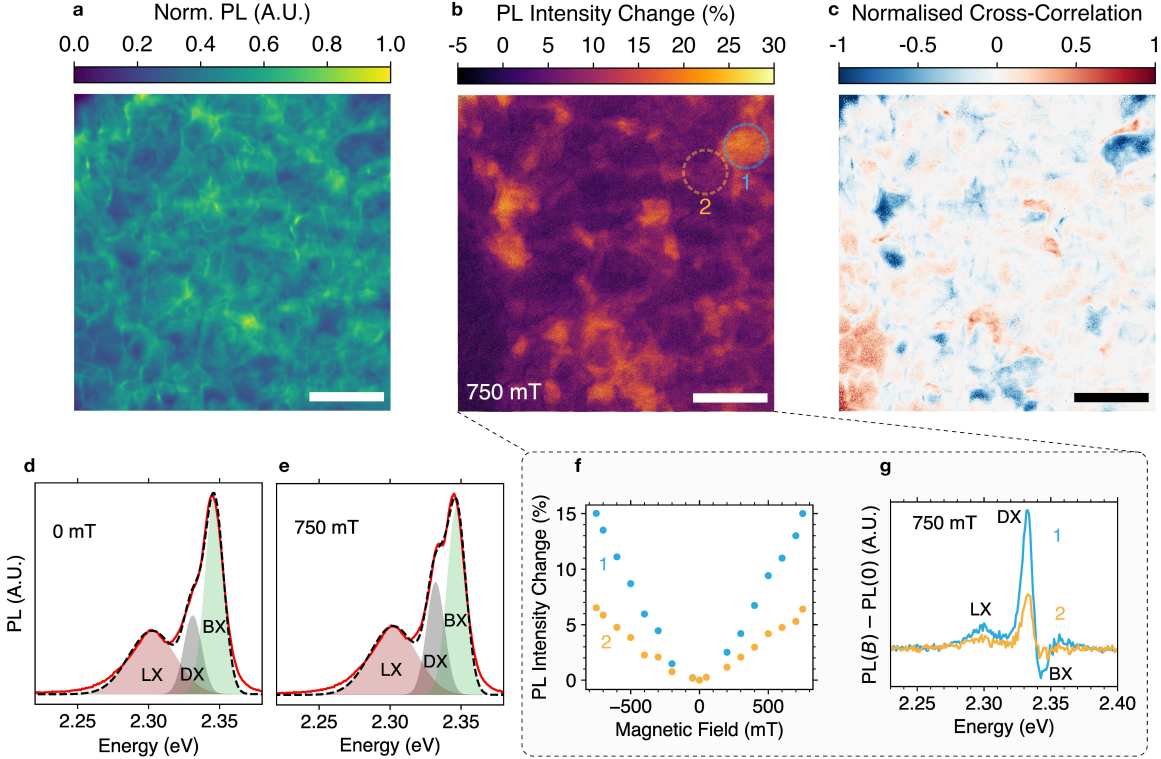

Figure 2: **Spatial heterogeneity of dark-exciton brightening in (PEA)$_2$PbI$_4$ thin films.** (a) PL microscopy image of (PEA)$_2$PbI$_4$ thin film at zero magnetic field at 3 K. (b) Spatial map of PL intensity change under 750 mT magnetic field ($\Delta$PL), obtained using Equation 1. (c) Correlation map showing cross-correlation between zero field and $\Delta$PL images. The scale bar in (a) to (c) is 50 µm. Decomposed PL spectrum measured (red) and fitted (black dashes) for entire illuminated area at (d) zero magnetic field and (e) 750 mT. Shaded regions show contributions to the fit from the LX (red), DX (grey), and BX transitions (green). (f) Change in PL intensity and (g) spectral change under 750 mT applied magnetic field for regions 1 and 2 labelled in (b).

As expected, we observed an increase in the overall PL intensity proportional to the applied magnetic field (Figure 2f) due to transfer of oscillator strength and funnelling of exciton population to the lower-energy DX state at low temperatures.[51] Pixel-wise $\Delta$PL at $B$

= 750 mT is shown in Figure 2b, unveiling dark-exciton emission at a high resolution without the need for spectrally resolving each pixel. Results in Figure 2b reveal that ΔPL is non-uniform across the sample region, highlighting the spatially inhomogeneous nature of dark-exciton brightening.

The features identified in the ΔPL image (Figure 2b) appear to partially resemble those seen in the zero-field PL image (Figure 2a) but have distinctive characteristics. To globally quantify these features, we calculated an average correlation length of $L_{\Delta PL}$ = 11.96 ± 0.44 µm for the amplitude of ΔPL by using a spatial autocorrelation analysis method (Experimental Section and Figure S3). Interestingly, this value is larger than that obtained for the zero-field image, which was determined to be $L_{PL(0)}$ = 7.35 ± 0.66 µm, suggesting longer range correlation of ΔPL compared to the zero-field PL. This discrepancy is likely a result of smaller correlated features of the zero-field PL (Figure 2a), such as the very bright grain boundaries acting to reduce $L_{PL(0)}$ which are not prominent in the ΔPL image (Figure 2b), in addition to large correlated areas of low ΔPL (Figure 2b) which increase the value of $L_{\Delta PL}$.

From Figure 2b, we see that regions with largest ΔPL typically correlate with regions of low intensity zero-field PL (Figure 2a), such as the area in the upper-right corner (labelled region '1' in Figure 2b, f, and g). However, this rule is not consistent across the sample, exemplified at the lower-left corner of the images, where darker zero-field PL (Figure 2a) has a smaller corresponding ΔPL (Figure 2b).

To quantify the relationship between the zero-field PL (Figure 2a) and ΔPL (Figure 2b) images, we calculated the cross-correlation between the two images (see Experimental Section) and the resulting correlation map is shown in Figure 2c. The regions with blue colour on the correlation map represent negative correlation and red areas represent positive correlation. This analysis reveals a spatially non-uniform mixture of positive, negative, and small correlation between zero-field PL and ΔPL. We calculated Pearson's correlation coefficient, *r* = −0.23 (see Experimental Section and Figure S4), indicating a small overall anticorrelation, as expected from inspection of Figure 2c.

One would expect that lower intensity regions in the zero-field image (Figure 2a) correspond to areas of localised dark-exciton emission (or reduced bright exciton emission), since the DX transition is optically forbidden and should be less efficient than other recombination processes. Additionally, the localised regions of high ΔPL should correlate directly to a larger local density of dark excitons, assuming the modification of oscillator strength occurs uniformly. Indeed, we found that the areas of largest ΔPL had a larger relative DX spectral component at zero field (Figure S5).

These considerations explain the negative correlation between the zero-field PL and ΔPL (Figure 2c), suggesting spatially separated regions with differing dark and bright exciton concentrations. Since the correlation lengths $L_{\Delta PL}$ and $L_{PL(0)}$ are comparable to the grain size and the PL (ΔPL) patterns clearly relate to morphological features, the localisation of excitons is likely a result of the polycrystallinity of the (PEA)$_2$PbI$_4$ thin film producing spatially varying levels of disorder.

Mixing of exciton states in 2D perovskites is known to depend on the orientation of the crystallographic plane with respect to the applied magnetic field, with the largest effect observed in the Voigt geometry.[42,63] Due to the polycrystalline nature of the film, different regions contain differently oriented crystals and hence experience differing amounts of mixing between bright and dark exciton states. Hence, partial vertical alignment of grains can explain the regions of unexpected positive correlation between the zero-field PL and ΔPL (Figure 2c). Overall, the spatial variation of ΔPL most likely originates from a combination of crystal orientation, exciton localisation, or self-absorption effects.

Finally, we identified dark regions of near-zero ΔPL at the grain boundaries (Figure 2b) which correspond to bright lines in the zero-field image (Figure 2a), shown as a negative correlation in Figure 2c. We propose that emission in these regions originates from the bright exciton with out-of-plane (*z*) dipole moment orientation, which is evidenced further in the following section.

## Polarisation-resolved imaging of (PEA)$_2$PbI$_4$ thin films

To further investigate the role of exciton fine structure in the spatial variation of PL in (PEA)$_2$PbI$_4$, we used polarisation-resolved measurements to decouple the contribution of different exciton states. The BX sub-levels with dipole moments oriented to the three directions of the crystal plane (*x, y, z*) couple to light with electric field (***E***) linearly polarised in the corresponding direction. [41,42,52,55,59,62] This principle is shown in Figure 3g, with each sub-level labelled as BX$_x$, BX$_y$, and BX$_z$. However, their exact energetic ordering is a subject of ongoing debate. [41,42,52,55,59,62]

Under an external magnetic field, two pairs of states are produced which can be expressed as linear combinations of the zero-field states (Supporting Note 2). The first pair ($\psi_\parallel^1$ and $\psi_\parallel^2$) contains the DX state mixed with the in-plane bright states only (BX$_x$ and BX$_y$) and couples to light with electric field parallel to the magnetic field (***E***∥***B***), depicted in Figure 3g. The remaining pair ($\psi_\perp^1$ and $\psi_\perp^2$) which includes the out-of-plane bright exciton (BX$_z$) couples to light with electric field perpendicular to the magnetic field (***E***⊥***B***). Therefore, we can image the local dipole moment orientation by repeating the experiments shown in Figure 1, while resolving the collected light into two linear polarisation orientations: perpendicular to the magnetic field (***E***⊥***B***) and parallel to the magnetic field (***E***∥***B***).

Figures 3a–d, S6-S7 show the results of the polarisation-resolved PL imaging under an external magnetic field, which is applied in the horizontal direction of the page. As expected, we obtain a larger ΔPL when the collected light is polarised parallel to the applied field (***E***∥***B***), due to DX (and mixed BX sub-levels) having a dipole moment oriented in the direction of the applied field (Figures 3b and c). [41,42,52,55,59,62]

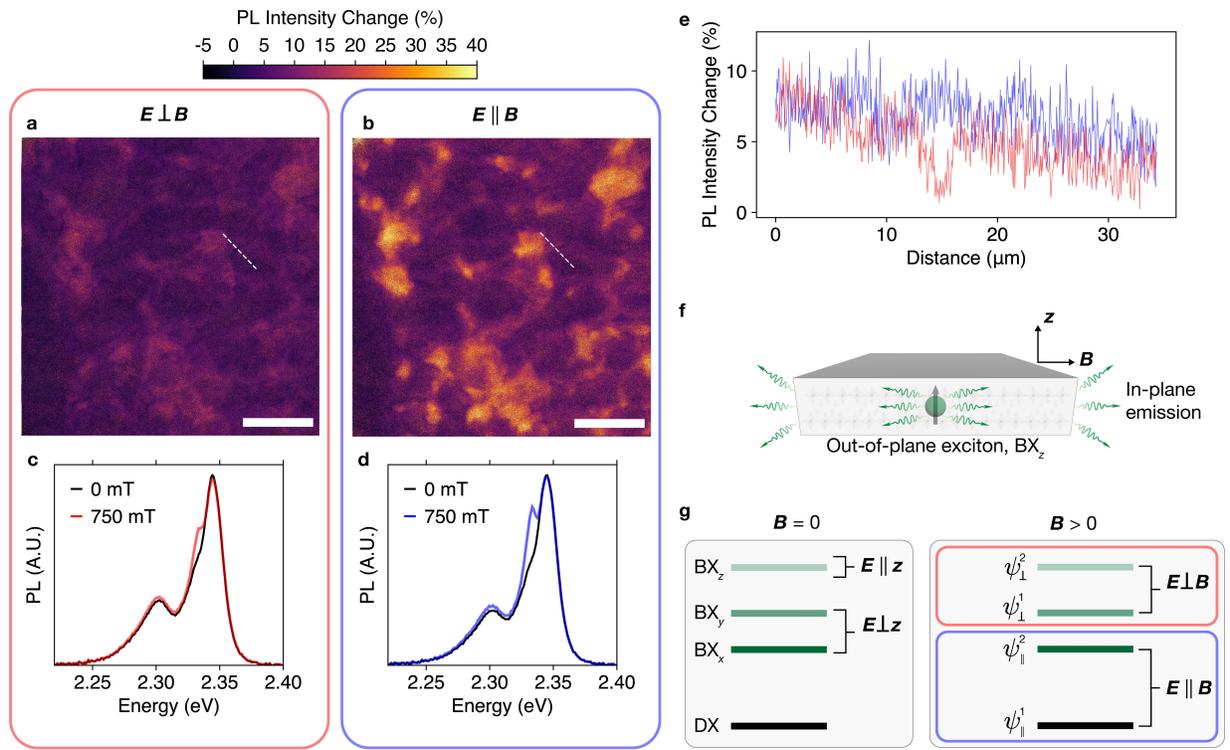

Figure 3: **Heterogeneous emission of exciton fine structure components in (PEA)$_2$PbI$_4$ thin films revealed by polarisation-resolved PL microscopy.** Spatial map of PL intensity change in (PEA)$_2$PbI$_4$ thin film at 3 K under 750 mT magnetic field (ΔPL) with linearly-polarised emission resolved (a) perpendicular to field (***E***⊥***B***) and (b) parallel to field (***E***∥***B***). PL spectra of entire collection area at zero field and 750 mT for (c) perpendicular polarisation (***E***⊥***B***) and (d) parallel polarisation (***E***∥***B***) with respect to magnetic field. (e) Line profile corresponding to dashed line in (a). The red line corresponds to perpendicular polarisation (***E***⊥***B***) and blue is for parallel polarisation (***E***∥***B***). (f) Schematic showing the out-of-plane bright exciton (BX$_z$) and its corresponding in-plane emission. (g) Schematic showing the exciton fine structure modification under magnetic field. The scale bar for all images is 50 µm.

Comparison of the polarisation-resolved ΔPL images (Figures 3a and b) reveal that the dark regions at grain boundaries (also seen in the non-polarisation-resolved image, Figure 2b) are primarily associated with light having perpendicular polarisation (***E***⊥***B***, Figure 3a). This is highlighted in Figure 3e, showing a line profile from each of the polarisation-resolved ΔPL images, with the location indicated by the white dashed line in Figures 3a and b.

We propose that emission at grain boundaries corresponds to PL originating from the out-of-plane BX state ($BX_z$), which has its Poynting vector in the direction parallel to the substrate plane (assuming the majority of grains grow with their 2D planes oriented parallel to the substrate).[59,60] As illustrated in Figure 3f, light emitted from this state will propagate parallel to the substrate and hence be waveguided to the edge of the grain, where it is then scattered and detected. Indeed, Posmyk et al. recently showed that the reflectance spectrum of the out-of-plane transition ($BX_z$) is exclusively observed from the edge of $(PEA)_2PbI_4$ single crystals.[62]

The waveguided in-plane emission from $BX_z$ corroborates with low ΔPL at the grain boundaries, since this state is the only BX sub-level which does not mix with the DX state in the Voigt configuration (Figure 3g).[41,55] The reduced ΔPL at grain boundaries is significantly more pronounced for the ***E⊥B*** configuration (Figure 3a) because this configuration selectively probes the pair of states which contain the $BX_z$ component ($\psi_\perp^1$ and $\psi_\perp^2$) and does not include the brightened DX state (Supporting Note 2). This observation is highlighted in the line profile shown in Figure 3e.

## Magneto-optical imaging of $(PEA)_2PbI_4$ single crystals

To further investigate the spatial variation of dark-exciton brightening in $(PEA)_2PbI_4$, we synthesised single crystals using a solution temperature lowering (STL) method (Figures S8 and S9) described in the Experimental Section and again performed the magneto-optical experiments outlined in Figure 1. Figure 4a shows a PL image at 3 K of $(PEA)_2PbI_4$ single crystal flakes with varying shapes and dimensions and Figure 4b shows the corresponding change in PL intensity under magnetic field (ΔPL), as given by Equation 1. The corresponding raw PL images are shown in Figure S10.

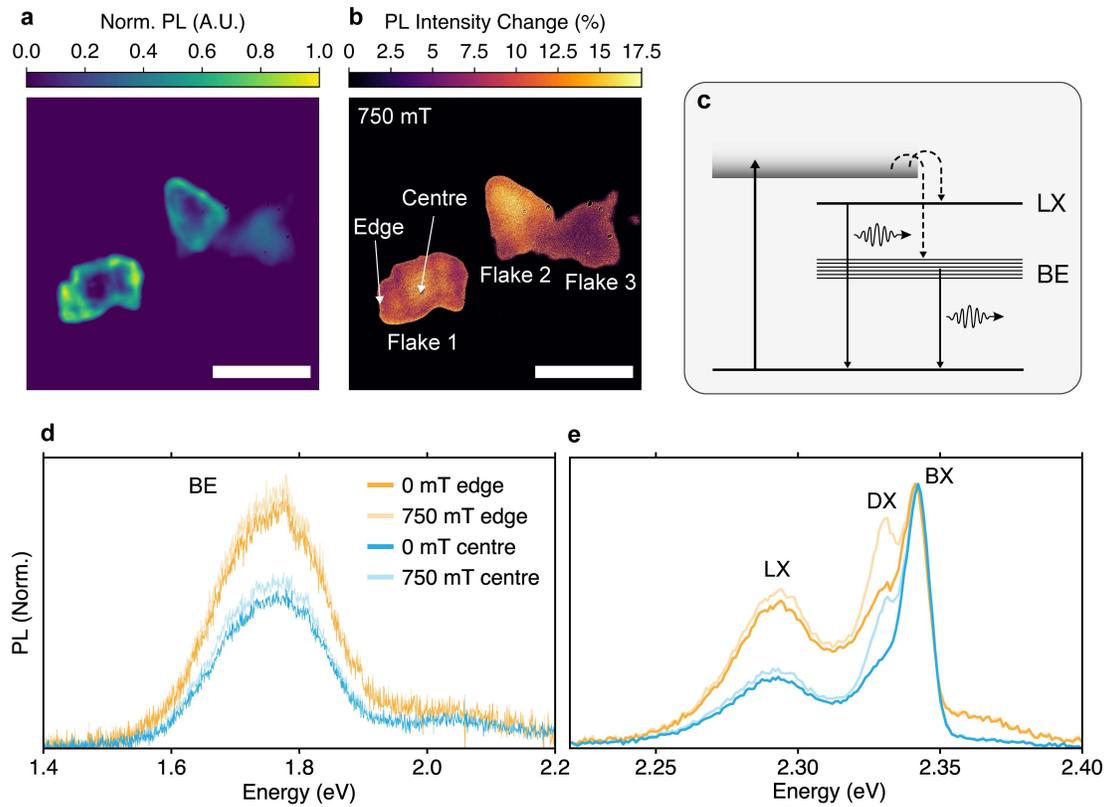

Figure 4: **Dark-exciton localisation at the edge of (PEA)$_2$PbI$_4$ single crystal flakes revealed by PL microscopy with an applied in-plane magnetic field.** (a) PL image at zero magnetic field of single crystal flakes at 3 K. (b) PL change under applied in-plane magnetic field of 750 mT (ΔPL). (c) Schematic showing the process of free excitons becoming trapped by either a shallow defect (LX) or mid-gap trap states resulting in broad emission (BE). (d) Lower-energy broad emission spectra obtained at zero field and 750 mT for the edge and centre of Flake 1. (e) Band-edge exciton PL spectra obtained at zero field and 750 mT for the edge and centre of Flake 1. All spectra are normalised to the BX peak. The y-axis for (d) has been enlarged compared to (e) to highlight spectral features. The scale bar for all images is 50 μm.

From the zero-field PL image (Figure 4a), again we observe a slight waveguiding effect producing brighter regions on the edge of the crystals, similar to the grain boundaries of thin films (Figure 2a) and previous reports on single crystals.[60,61] This is highly dependent on the shape and thickness of the crystal, since this will impact the critical angle of coupled light and influence total internal reflection (Figures S8 and S9). Crystalline flakes with more rounded edge morphology were chosen (Figure S9c–f) to distinguish them from the plate-like grains in the polycrystalline sample (Figure 2), suppressing the waveguiding effect to produce a more accurate representation of the localised emission.

Figures 4d and e show PL spectra at zero field and 750 mT for emission collected at the centre and at the edge of Flake 1 (Figure 4b), normalised to the BX emission. Similar to the behaviour of thin films (Figure 2d-g), the DX and LX emission increases with applied field in (Figure S11c). As commonly observed in 2D perovskite single crystals, there is also a broad lower energy emission, which has been proposed to originate from a combination of self-trapped excitons and excitons trapped by extrinsic defects.[64] However, the precise origins of this emission have been the topic of debate,[64–68] and therefore we simply refer to this spectral feature as the broad emission (BE). Whether the origin of this emission is self-trapping, extrinsic defects, or a combination, it is clear that trapped excitons occupy a broad range of states which extend to deep within the bandgap.

Interestingly, we find that BE increases slightly under an applied magnetic field, following the same trend as the LX emission (Figures S11c and d). The increase with magnetic field is significantly lower compared to the trend of DX, suggesting that BE and LX populations are comprised of excitons with a mixture of bright ($J = 1$) and dark ($J = 0$) character. The proportion depends on the Boltzmann distribution of excitons, with preferential trapping of dark excitons expected since DX is the lowest lying exciton state in the fine structure manifold and has been shown to be have a larger population at low temperature.[51] This finding proves that LX and BE are not part of the fine structure manifold and are a result of exciton trapping.

We found that spectra acquired from the edge of crystal flakes have a substantially larger relative contribution from DX, LX, and BE (lower relative BX emission) when compared to the centre (Figures 4d and e). This is repeatable for different flakes, and the corresponding spectra for Flake 2 can be found in Figure S11a and b. Therefore, our results suggest that photoexcited excitons can be bound by either shallow (LX) or deep defects (BE) which are concentrated at edges in the case of $(PEA)_2PbI_4$ single crystals. Following this, they can undergo PL, as summarised in Figure 4c. We note that the LX emission has also been previously attributed to biexcitons but remains controversial.[57,61,69,70]

The larger relative spectral component of DX at the edge of crystals opposes the results from thin films (Figure 2b), where almost no DX brightening is observed at grain boundaries. These differences are likely due to waveguided edge-emission of $BX_z$ being dominant in films compared to single crystals, since in the single crystals light must cover a larger distance to reach the edges.

These results unveil crucial distinctions between thin films and single crystals regarding exciton localisation. However, the spatial heterogeneity of the emission spectra in 2D perovskites requires further investigation to provide a complete picture of photophysical processes in these materials.

# Conclusion

This work represents the first study of perovskite materials using magneto-optical microscopy, revealing separate regions of localised dark and bright exciton populations in polycrystalline thin films of the 2D perovskite $(PEA)_2PbI_4$. Such localisation is anticipated to be beneficial for light emitting applications where a high spatial density of excitons is required, such as lasers. Using polarisation-resolved measurements, we have identified directionality of emission from excitons with an out-of-plane dipole moment in 2D perovskite thin films, which is a vital factor for outcoupling of light in optoelectronic devices. Moreover, this effect could be harnessed for applications involving the selectivity of exciton states such as spintronics and spin-orbitronics.

In single crystals of $(PEA)_2PbI_4$, we have shown that excitons can become trapped at the edge of crystals by either a shallow defect state (LX) or a broad distribution of mid-gap states (BE). The discrepancy between recombination processes in single crystals and thin films of $(PEA)_2PbI_4$ is a key consideration when selecting designs for optoelectronic technologies.

In summary, this work has demonstrated magneto-optical microscopy as a powerful technique to probe exciton emission in RP phase 2D perovskites. Our findings provide detailed insight into the heterogeneous light-emitting properties of these materials, which have major implications for the scalability of optoelectronic devices and advancing new technologies which leverage the manipulation of exciton states. If the contributions to spatially dependent magnetic field effects can be successfully decoupled, magneto-optical microscopy could be utilised for mapping the orientation of grains in RP phase 2D perovskite thin films which is crucial for electronic transport and should be the subject of future investigations.

# Experimental Section

*Materials*: Lead iodide (PbI$_2$, 99.999%) was purchased from TCI. N,N-dimethylformamide (DMF, 99.99%), dimethyl sulfoxide (DMSO, 99.50%), lead oxide (PbO), and hydroiodic acid (HI, 57% aqueous solution) were purchased from Sigma-Aldrich. Phenethylammonium Iodide (PEAI) was purchased from Xi'an Polymer Light Technology Corporation.

*(PEA)$_2$PbI$_4$ Thin-Film Fabrication*: Thin films of (PEA)$_2$PbI$_4$ were deposited on sapphire substrates from precursor solutions prepared by dissolving the mixture of 2:1 (PEAI:PbI$_2$ in DMF:DMSO=4:1 (v:v). The 2D perovskite layers were deposited on the substrates by spin-coating at 5000 rpm for 30 s, with an acceleration of 4000 rpm/s. Following this, the coated substrates were transferred onto a hotplate and annealed at 100 °C for 5 mins. X-ray diffractogram of a (PEA)$_2$PbI$_4$ thin film can be found in Figure S12.

*(PEA)$_2$PbI$_4$ Single Crystal Synthesis*: Millimetre-sized single crystals of (PEA)$_2$PbI$_4$ were synthesised using a solution temperature lowering (STL) method. Lead oxide and PEAI with a concentration of 1.72 and 3.45 mmol, respectively, was added to HI. The temperature was increased to ≈ 90°C, and then cooled by 1°C per hour to obtain crystals suspended in solution. We then carried out an exfoliation technique by adding the crystals to an antisolvent (chlorobenzene) and sonicating for 5 mins to obtain smaller flakes on the scale of 10–100 µm which were suitable for microscopy analysis. X-ray diffractogram of a (PEA)$_2$PbI$_4$ single crystal can be found in Figure S12.

*Photoluminescence Microscopy*: The photoluminescence was measured using a microscopy system housed in a Montana Instruments Magneto-Optic cryostat module at 3K, with the sample held in a vacuum chamber. The excitation was provided by a 405 nm Thorlabs M405L2 LED which was directed through a lens before the microscope objective in order to produce illumination with a uniform flat-top profile. The PL was obtained through the same objective lens as the excitation and then passed through a 450 nm long-pass filter.

The PL was then focused onto a QHY68M camera for spatially resolved measurements or a Thorlabs CS200 fibre-coupled spectrometer for spectrally resolved measurements. To account for potential temporal changes in the sample PL due to degradation for example, the in-plane magnetic field was varied using randomised field values. Additionally, for the spectrally resolved measurements the field was alternated between zero and non-zero values to calculate the PL intensity change using equation 1.

*Autocorrelation Function Analysis*: To quantify spatial correlation for the magnetooptical microscopy images, we performed 2D autocorrelation function (ACF) analysis using a fast Fourier transform (FFT)-based approach. Images were binned over regions of 16×16 pixels (≈1×1 µm) to eliminate effects from noise and the mean intensity was subtracted to centre the data. The image $I(x,y)$ was then transformed to Fourier space using the FFT, and the power spectrum was calculated as the squared magnitude of the FFT output:

$$P(f_x, f_y) = |\mathcal{F}\{I(x,y)\}|^2 \qquad (2)$$

where $\mathcal{F}$ denotes the Fourier transform, and $f_x, f_y$ are the spatial frequency components. The 2D autocorrelation function $A(x,y)$ was obtained by applying the inverse FFT to the power spectrum:

$$A(x,y) = \mathcal{F}^{-1}\{P(f_x, f_y)\} \qquad (3)$$

The autocorrelation result was then shifted to centre the zero-lag position and normalised to its maximum value. The resulting ACF provides a spatial map of correlations, from which correlation lengths can be extracted.

We computed the radial average of the autocorrelation function to provide an isotropic estimate of the correlation length. The radial profile was obtained by averaging the autocorrelation values over concentric circles around the ACF centre. The radial distance *r* at each pixel was computed as:

$$r = \sqrt{(x - x_0)^2 + (y - y_0)^2} \tag{4}$$

where $x_0, y_0$ are the coordinates of the ACF centre. The radial profile was then calculated by binning the ACF values based on integer values of *r*. The correlation length was determined as the radial distance where the autocorrelation dropped to $1/e$ of its maximum value. A slight dependence of the correlation length on the bin size used was determined, so the mean values across a 16–32-pixel (≈1–2 µm) range of bin sizes was calculated and the standard deviation was given as the error.

*Cross-correlation of zero field PL and ΔPL images*: The cross-correlation between the zero-field PL image and the ΔPL image was calculated by first performing Z-score normalisation. This is achieved by using the non-normalised intensity at each pixel, subtracting the mean intensity across the image, and dividing by the standard deviation of intensity, given by,

$$I_{\text{norm}} = \frac{I - \langle I \rangle}{\sigma_I} \tag{5}$$

where *I* is the intensity, $\langle I \rangle$ is the mean intensity, and $\sigma_I$ is the standard deviation. Pixel-wise multiplication was then performed between the normalised images to generate a correlation map.

*Pearson's r correlation*: The Pearson correlation coefficient (*r*) between the zero-field PL image and the PL intensity change image was calculated to quantify the linear relationship. First, Z-score normalization was applied to both images using equation 2. The normalized images were then binned using a bin size of 50 × 50 pixels (3.12×3.12 µm). The binned images were subsequently flattened, and Pearson's correlation coefficient was computed using:

$$r = \frac{\sum(x_i - \bar{x})(y_i - \bar{y})}{\sqrt{\sum(x_i - \bar{x})^2 \sum(y_i - \bar{y})^2}} \tag{6}$$

where $x_i$ and $y_i$ are the binned intensity values from the zero-field and difference images, respectively, and $\bar{x}$ and $\bar{y}$ are their means.

## Author contributions

C.G.B., A.M, N.P.S., D.R.McCamey, and A. H.-B. contributed to the conceptualisation. C.G.B. contributed to the literature review, writing of the original draft, synthesis of perovskite thin films, magneto-optical measurements, and data analysis. A.M. developed the magneto-optical microscopy setup, assisted with the measurements using the setup, and assisted with data analysis. T.L.L. and C.L. synthesised the perovskite single crystals. N.P.S. performed the XRD and SEM measurements, and assisted with data analysis. The supervision, project administration, funding acquisition, and resources were provided by A.H.-B., and D.R.McCamey. D.R.McKenzie assisted with the supervision. All authors assisted to the review and editing of manuscript.

## Acknowledgement

This work is supported by the Australian Government via the Australian Research Council (ARC) Centre of Excellence in Exciton Science (CE170100026) for D.R. McCamey, A.M., and N.P.S., ARC Future Fellowship FT210100210 for A H.-B.; and the Australian Renewable Energy Agency (ARENA) for C.G.B., T. L. L., C. L., D. R McKenzie, and A H.-B.. N.P.S. acknowledges the support from an Australian Government Research Training Program (RTP) Scholarship. T. L. L. is supported by the University of Sydney Faculty of Science Postgraduate Research Excellence Award and C. L. was supported by the John Hooke Chair of Nanoscience Postgraduate Research Scholarship. The authors acknowledge the use of facilities in the Solid State and Elemental Analysis Unit at Mark Wainwright Analytical Centre at the University of

# Supporting Information for 'Revealing localised dark exciton populations in 2D perovskites via magneto-optical microscopy'

Christopher G. Bailey,[*,1] Adrian Mena,[2] Tik Lun-Leung,[1] Nicholas Sloane,[2] Chwenhaw Liao,[1] David R. McKenzie,[1] Dane R. McCamey,[*,2] and Anita Ho-Baillie[*,1]

[1]*School of Physics, University of Sydney, Sydney, Australia*

[2]*ARC Centre of Excellence in Exciton Science, School of Physics, University of New South Wales, Sydney, Australia*

* E-mail: Christopher.Bailey@sydney.edu.au; Dane.McCamey@unsw.edu.au; Anita.Ho-Baillie@sydney.edu.au

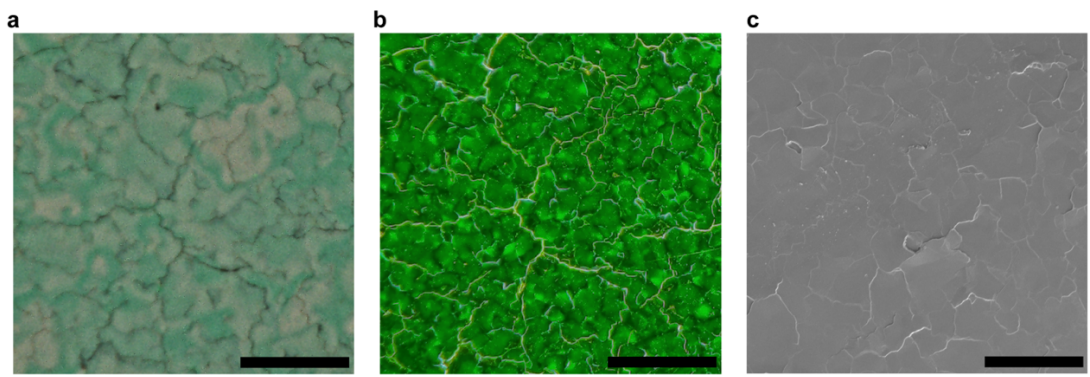

Figure S1: **Morphology of (PEA)$_2$PbI$_4$ thin films.** (a) Bright field optical microscope, (b) dark field optical microscope, and (c) scanning electron microscope (SEM) image of (PEA)$_2$PbI$_4$ thin films. The scale bar for all images is 50 μm.

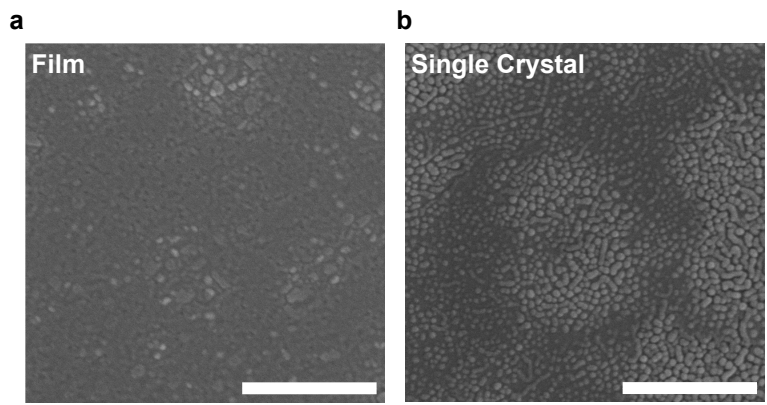

Figure S2: **Nanoscale texturing of (PEA)$_2$PbI$_4$ thin films and single crystals.** SEM image of nanoscale texturing on (a) (PEA)$_2$PbI$_4$ thin film and (b) single crystal. The scale bar for all images is 500 nm.

## Supporting Note 1

Figure S1 shows results of bright- and dark- field optical microscopies, and scanning electron microscopy (SEM) of (PEA)2PbI4 thin films. Plate-like flat regions with widths ranging from ≈ 5–40 μm are identified as grains. Smaller features on a scale of 5–50 nm observed under SEM for thin films (Figure S2a) and single crystals (Figure S2b) are associated with texturing of the surface rather than crystalline domains. One possible origin of this texturing could be vacuum- or electron-beam- induced degradation during SEM experiments.[1,2]

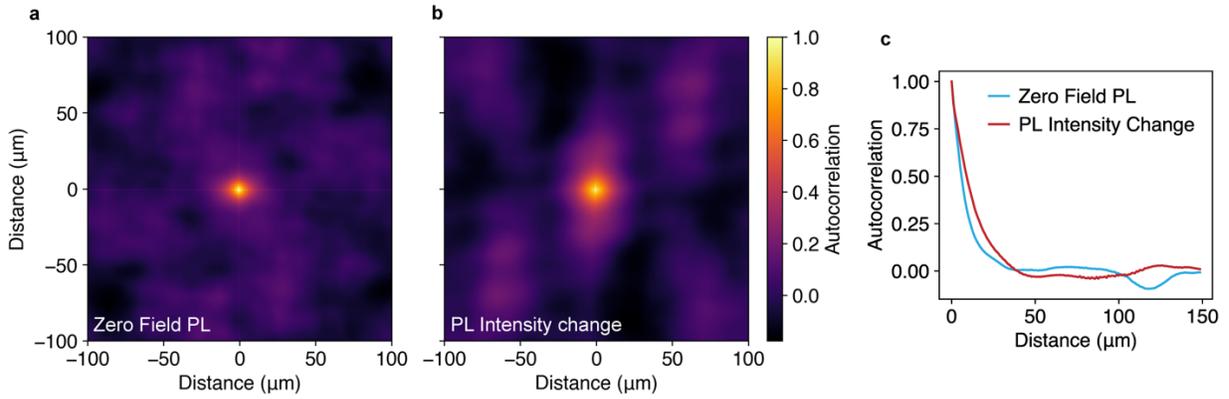

Figure S3: **Spatial autocorrelation analysis of magneto-optical microscopy images.** 2D map of the autocorrelation function for intensities of (a) the zero-field PL image and (b) the PL intensity change image (ΔPL) for $(PEA)_2PbI_4$ thin films. (c) Mean radial profiles of the autocorrelation for the zero-field PL and intensity change images. Images were binned over regions of 16×16 pixels (≈1×1 μm).

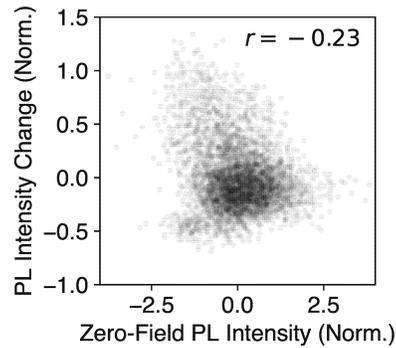

Figure S4: **Pearson's correlation coefficient for correlation between zero-field PL and PL intensity change images.** Pearson's correlation coefficient for correlation between intensities of zero field PL image and spatial map of PL intensity change (ΔPL) under applied magnetic field of 750 mT for $(PEA)_2PbI_4$ thin films. Images were binned over regions of 50×50 pixels (3.12×3.12 μm).

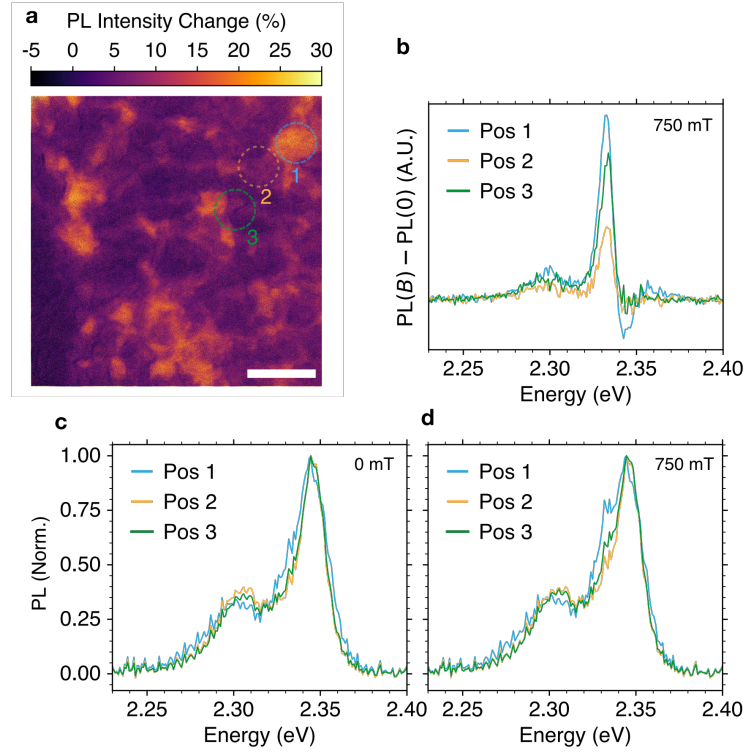

Figure S5: **Low temperature PL microscopy of (PEA)$_2$PbI$_4$ thin films with an applied in-plane magnetic field.** (a) Spatial intensity map (scale bar = 50 μm.) and (b) spectral change of PL intensity in (PEA)$_2$PbI$_4$ thin film at 3 K under 750 mT magnetic field (ΔPL). PL spectra of different regions under (c) zero and (d) 750 mT magnetic field.

## Supporting Note 2

Under an external magnetic field, two pairs of exciton states are formed with one pair ($\psi_\parallel^1$ and $\psi_\parallel^2$) that couples to light polarised with **E** parallel to **B** and another pair ($\psi_\perp^1$ and $\psi_\perp^2$) which couples to light polarised with **E** perpendicular to **B**. These states which can be expressed as a linear combination of the zero-field states as follows:

$$\psi_\parallel^1 = c_1 \text{DX} + d_1(\text{BX}_x - \text{BX}_y) \tag{1}$$

$$\psi_\parallel^2 = c_2 \text{DX} + d_2(\text{BX}_x - \text{BX}_y) \tag{2}$$

$$\psi_\perp{}^1 = c_3 BX_z + d_3(BX_x + BX_y) \tag{3}$$

$$\psi_\perp{}^2 = c_4 BX_z + d_4(BX_x + BX_y), \tag{4}$$

where $c_1$–$c_4$ and $d_1$–$d_4$ are coefficients described elsewhere.[4,5]

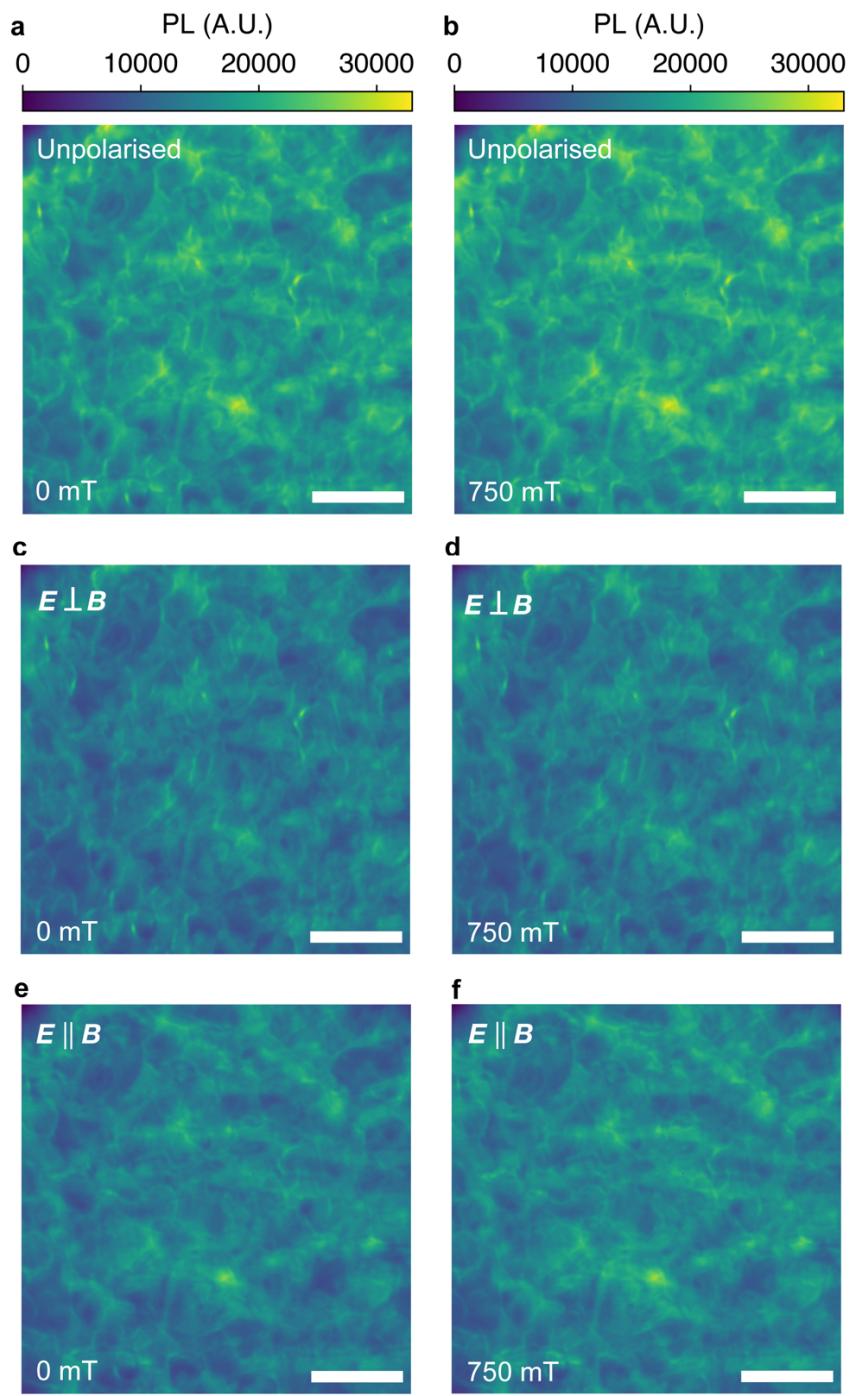

Figure S6: **Raw images of low temperature PL microscopy of (PEA)$_2$PbI$_4$ thin films with an applied in-plane magnetic field.** PL microscopy image of (PEA)$_2$PbI$_4$ thin film at 3 K with (a) zero magnetic field and (b) 750 mT for non-polarised emission. Corresponding PL microscopy images for perpendicular polarisation (c), (d), and parallel polarisation (e), (f). The scale bar for all images is 50 μm.

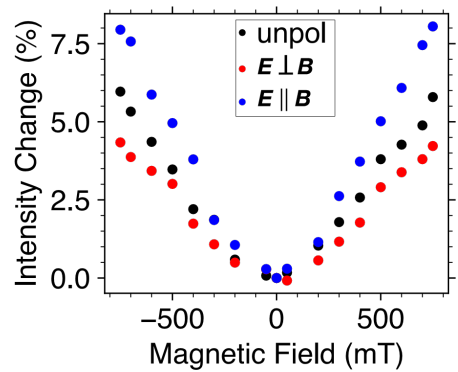

Figure S7: **Total integrated PL intensity change of (PEA)$_2$PbI$_4$ thin film under applied magnetic field.** PL intensity change (ΔPL) of (PEA)$_2$PbI$_4$ thin film under applied magnetic field, integrated over the entire image area for non-polarisation-resolved (black), for polarisation perpendicular to the magnetic field (red), and for polarisation parallel to the magnetic field (blue).

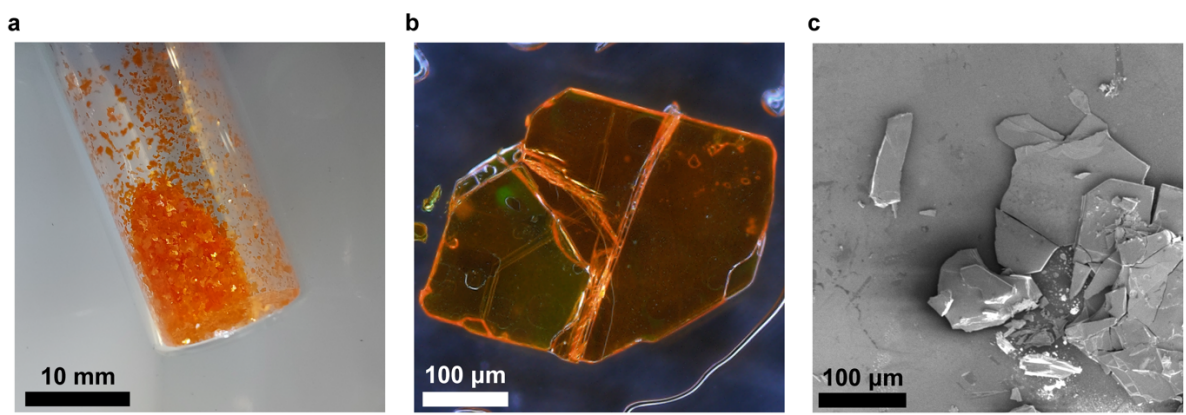

Figure S8: **As-grown single crystals of (PEA)$_2$PbI$_4$.** Image of (PEA)$_2$PbI$_4$ crystals in (a) vial, with typical size of approximately 0.5–1 mm. (b) Dark-field microscope image of crystal flake. (c) SEM image of crystal flakes.

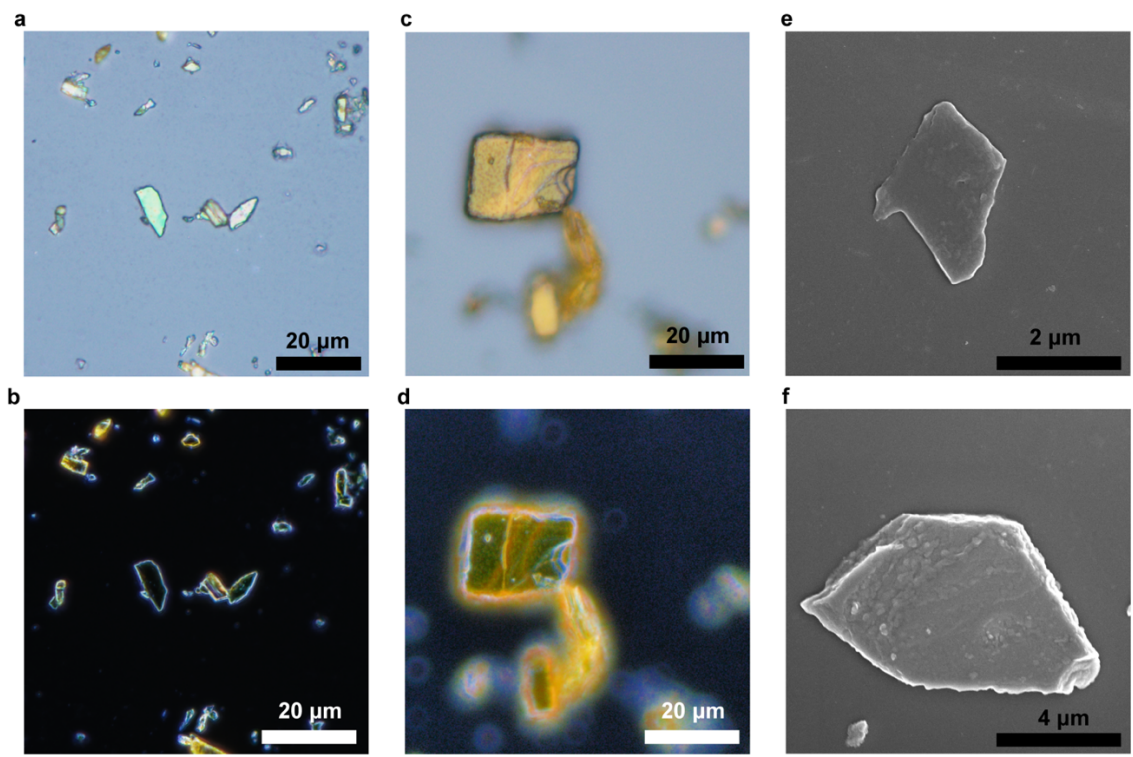

Figure S9: **Microscopy of exfoliated (PEA)$_2$PbI$_4$ single crystal flakes.** Bright-field (a), (c) and dark-field (b), (d) microscopy images of exfoliated flakes of (PEA)$_2$PbI$_4$. (e), (f), SEM images of exfoliated flakes of (PEA)$_2$PbI$_4$.

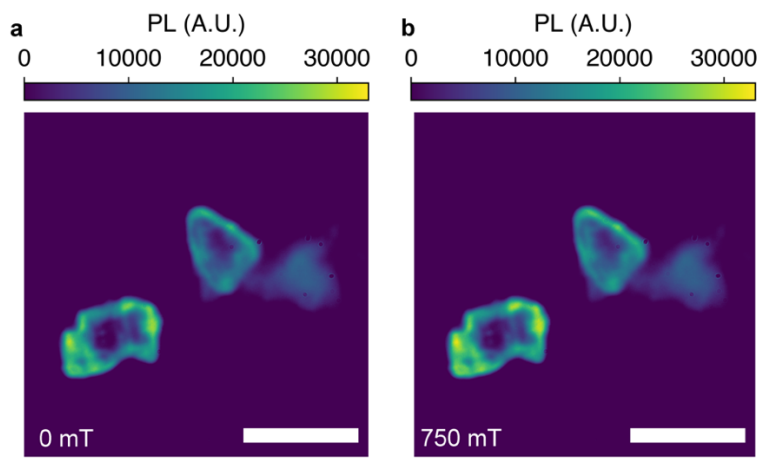

Figure S10: **PL microscopy of (PEA)$_2$PbI$_4$ single crystals with an applied in-plane magnetic field** at (a) zero magnetic field and (b) 750 mT. The scale bar for all images is 50 μm.

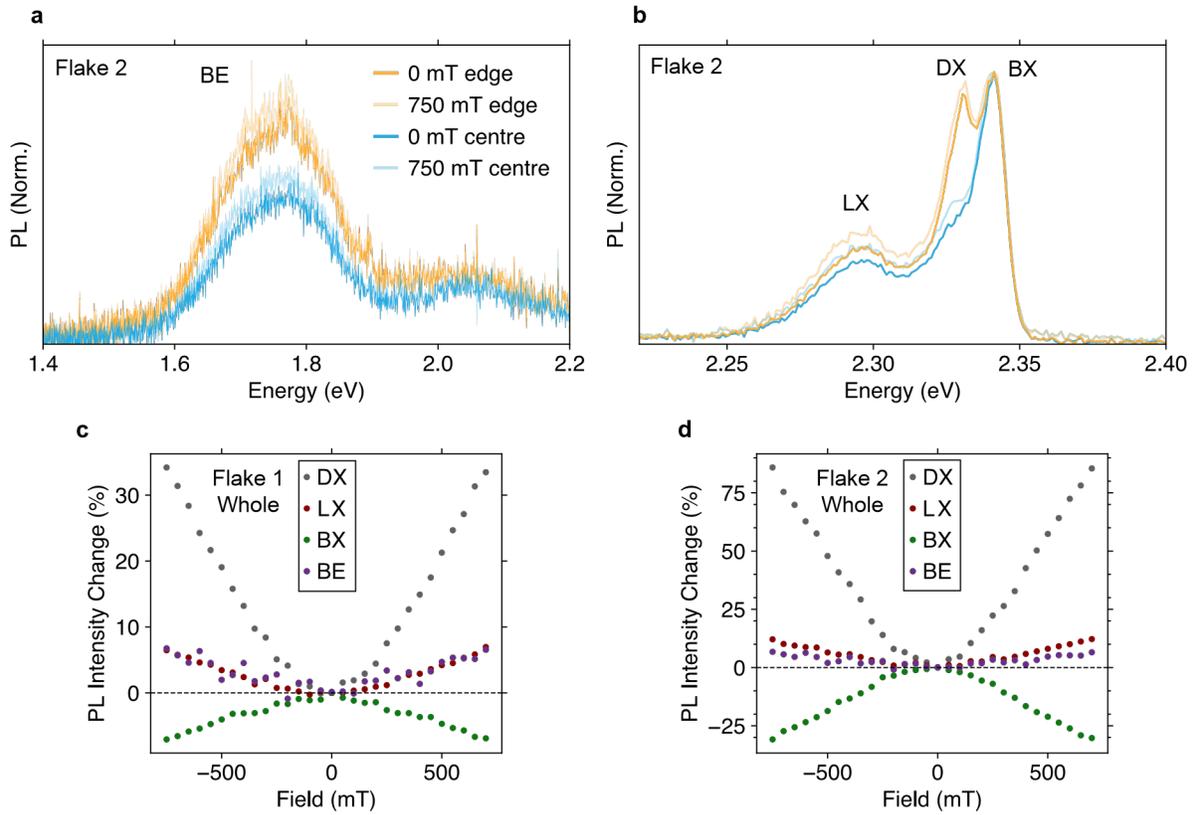

Figure S11: **Low temperature PL spectroscopy of $(PEA)_2PbI_4$ single crystal flake with an applied in-plane magnetic field.** (a) Lower-energy broad emission (BE) spectra obtained at zero field and 750 mT for the edge and centre of flake 2 (labelled in main text). (b) Band-edge exciton PL spectra obtained at zero field and 750 mT for the edge and centre of flake 2. Change in integrated PL intensity under magnetic field (ΔPL) for Gaussian fits to PL spectra obtained for the entire area of flake (c) 1 and (d) 2. All spectra are normalised to the BX peak.

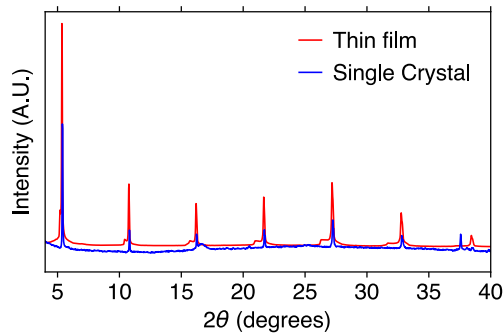

Figure S12: **X-ray diffractograms of a $(PEA)_2PbI_4$ thin film and single crystal.**